# Application of Artificial Neural Networks for Catalysis

Zhiqiang Liu, Wentao Zhou

East China University of Science and Technology, 200237 Shanghai, China

**Abstract**：Catalyst, as an important material, plays a crucial role in the development of chemical industry. By improving the performance of the catalyst, the economic benefit can be greatly improved. Artificial neural network (ANN), as one of the most popular machine learning algorithms, relies on its good ability of nonlinear transformation, parallel processing, self-learning, self-adaptation and good associative memory, has been widely applied to various areas. Through the optimization of catalyst by ANN, the consumption of time and resources can be greatly reduced and greater economic benefits can be obtained. In this review, we show how this powerful technique helps people address the highly complicated problems and accelerate the progress of the catalysis community.

# 1 Introduction

In recent decades, Artificial neural network (ANN) , as a non-linear fitting algorithm, has become one of the most popular machine learning techniques due to its advantages of easy-training, adaptive structure, and tunable training parameters[1,2]. With the development of algorithms, there are currently a large number of ANN methods, such as the back-propagation neural network (BPNN)[3,4], general regression neural network (GRNN) [5], and extreme learning machine (ELM)[6]. More recently, the deep neural network (DNN) has raised broad interests due to its strong learning capacity and the popular concept of deep learning techniques. Previous studies have shown that different neural network algorithms have different advantages for practical applications.

Nevertheless, to the best of our knowledge, there are very few review studies that summarize the ANN applications for catalysis research. Therefore, detailed discussions on the potential applications of ANNs for both experimental and theoretical catalysis are also necessary. Here, we aim to summarize the recent progress of ANN applications for catalysis, from the perspectives of the experiments.

# 2 Artificial neural networks

## 2.1 Overview of artificial neural networks

Artificial Neural Network, which is abbreviated as Neural Network, is an abstraction and simulation of the human brain. It is an interdisciplinary discipline involving biology, computer science, and mathematics. It is widely used in artificial intelligence and machine learning. It abstracts the neural network of human brain from the perspective of information processing, establishes some simple model, and forms different networks according to different connection modes. Artificial neural network is a kind of parallel interconnected network with adaptive self-learning adjustment, which is composed of the most basic unit group neurons. Through training and learning, artificial neural networks can simulate the biological nervous system to respond to specific objects, and its basic unit neuron is a simplification and simulation of biological neurons. Artificial neural networks are based on these simplifications and simulations of biological neurons[7].

The network system of artificial neural network is very complex, it is composed of many and single basic neurons, the neurons simulate the human brain to process information, and connect with each other, carrying on the nonlinear change to process information[8]. By training the information sample, the artificial neural network information processing information is input into the neural network, so that it has the memory and recognition function of human brain, and all kinds of information processing is completed. Artificial neural network has good non-linear transformation ability, parallel processing ability, self-learning and self-adaptation ability and good associative memory ability, but also avoid complex mathematical derivation, to ensure that the sample defect and parameter drift can ensure stable output[9].

As an important part of artificial intelligence, artificial neural network has the advantages of super robustness, fault tolerance, full approximation of any complex nonlinear relationship, parallel processing, learning and self-adaptation. There is a broad space for development in many fields involving the processing of nonlinear and complex problems. The main application areas are auxiliary assembly system [10,11], intelligent driving [12,13], Chemical product development [14-16], auxiliary medical diagnosis[17-19], image processing [20-22], automatic control of power systems [23-25], signal processing [26-28], process control and optimization[29-31], troubleshooting [32,33], game theory [34,35], etc.

## 2.2 Classification

Artificial neural network can be divided into feedback network and forward network in terms of structure, and can be divided into random network or deterministic network in terms of performance, which can also be called discrete network and continuous network. It can also be divided into management network and free network according to the method of learning. According to the nature of connection, it can be divided into first order linear correlation network and high order nonlinear correlation network. This paper focuses on the analysis of the topology of artificial neural network[36].

(1) Feedback network: Feedback network mainly includes BAM, Hamming, Hopfield, etc. Feedback networks with feedback between neurons can be represented by an undirected complete graph. The state of the neural network in the aspect of information processing is transformative and can be processed by using the dynamic system theory. The associative skill function of the system is closely related to the stability of the system, and Boltzmann machine and Hopfield network belong to this type.

(2) Forward network includes BP, multi-layer perceptron, single-layer perceptron, adaptive linear network, etc. In a forward network, each neuron in the network receives input from the previous level and outputs it to the next level. The network can be represented by a directed acyclic graph, which has no feedback. The network converts signals from the input space to the output space, and the multiple combinations of its information processing capabilities are derived from simple nonlinear functions. The network structure is easy to implement and relatively simple. Back propagation network is a typical forward network[37].

**2.3 Artificial neural network learning rules**

The learning rules of artificial neural networks are actually a way of network training. The purpose is to modify the weights of neural networks and adjust the thresholds of neural networks so that they can better complete some specific tasks. At present, neural networks have two different learning methods: tutored learning (also called supervised learning) and unsupervised learning (also called autonomous learning).

**2.3.1 Supervised learning**

The so-called supervised learning is a process in which the neural network requires training data to be supervised during the training process. This process is a process of continuously adjusting the weight under the effect of the expected output, that is, when the training data is input into the neural network after training [38]. After learning the output, the network compares the output with the expected output. If the output of the neural network is within the allowable range of errors compared to the expected output, then the neural network learning can be considered to have been completed. If it is not within the allowable range of errors, the neural network must continuously adjust the weights set to reduce the error, so that the output of the neural network is closer to the expected output, until the error is within the range allowed by the error, and the training ends. Therefore, it can be seen that the learning process with a mentor is a process of weight adjustment under the expectation of supervision. In this process, the change of the weight of the neural network reflects the learning process of the entire network. The final adjusted weight is this nerve. In this way, after continuous supervised learning, a neural network model with preliminary intelligence has been basically established[7,39].

### 2.3.2 Unsupervised learning

The difference between unsupervised learning, also known as autonomous learning and supervised learning, is that unsupervised learning does not have an external supervision mechanism. It has no expected output. The training data is not included in the output after being input into the network by the input layer. The entire neural network checks the characteristics and rules of the training input data, and formulates a judgment standard[1]. The network refers to this standard to adjust the weight. This kind of unsupervised learning can be considered as a kind of self-organized learning. The discrimination criteria formulated before training are also pre-set rules such as competition rules. Through the cooperation between the neurons, the network weights are continuously adjusted to respond to the input mode excitation until the entire neural network forms an ordered state.

### 2.4 Schematic Structure of an ANN

A complete algorithmic structure of a conventional ANN consists of at least three different layers: the input, hidden, and output layers (Figure 1). Each layer consists of a certain number of neurons. Each neuron inter-connects with all the neurons in the following layer. Each connection represents a weight that contributes to the fitting. With a proper activation function, a combination of optimized weights can generate the prediction of the dependent variable:

$$NET = \sum_{i,j}^{n} w_{i,j} x_i + b \qquad (1)$$

Where $w_{i,j}$ represents the weight value of a connection, $x_i$ represents an inputted independent variable, and b represents a bias. For the activation function ($f$ (NET)), the sigmoid function is one of the most popular forms that can introduce a smooth non-linear fitting to the training of an ANN (Equation (2)). The training of an ANN is essentially the optimization of each weight contribution based on the data groups in the training set. The most commonly used weight optimization method is the back-propagation algorithm, which iteratively analyzes the errors and optimizes each weight value based on the errors generated by the next layer. As we have mentioned above, there are also some other types of networks like GRNN and ELM. Though there are some differences in the weight training and algorithmic structures, the basic principles, as well as the training and testing processes, are very similar. More details about their principles can be found in References[38].

$$f(NET) = \frac{1}{1+e^{-NET}} \tag{2}$$

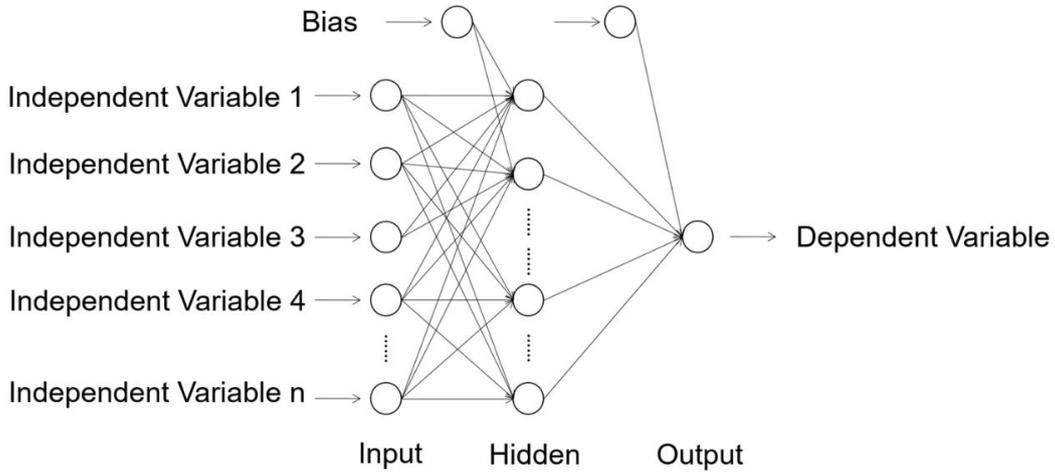

**Figure 1.** Algorithmic structure of a typical artificial neural network (ANN).

**2.5 Model Development**

The rational development of a knowledge-based ANN model consists of two parts: (i) training and (ii) testing. The training process is the so-called "learning" process from the database, while the testing process is the validation of the trained model using the data groups that are not previously involved in the training process. Detailed discussions are shown in Sections 2.5.1 and 2.5.2. It should be noted that the training and testing processes are not only suitable for the topic discussed in this review, but are also suitable for ANN model development in many other areas.

**2.5.1 Model Training**

The training of an ANN consists of database preparation and the selection of variables. The size of the database should be sufficiently large to avoid over-fitting. For each variable (especially the dependent variable), the data range should be wide enough to ensure good training. If the data range is too narrow, the trained model might only have a good prediction capacity in a very local region. For the numerical prediction, the dependent variables are usually the properties that are hard to acquire from regular measurements or calculations. The independent variables, on the other hand, should be easily-measured and have potential relationships with the selected dependent variable. More details about the training criterion can be found in Reference[40].

### 2.5.2 Model Testing

To validate the trained ANN, a testing process is necessary. The testing of a model should use the data groups that are not used for the training process. With the inputs of the testing set, the outputted data can be compared with the actual data in the testing set, with the root mean square error (RMSE) calculated by Equation (3):

$$RMSE = \sqrt{\frac{\sum_{i}^{n}(P_i - A_i)^2}{n}} \tag{3}$$

where $P_i$ represents the predicted value outputted by the ANN, $A_i$ is the actual value, and $n$ represents the total number of samples. If the calculated RMSE from the testing set is relatively small, it means that the ANN is well-trained. It should be noted that for the training and testing of an ANN, a cross-validation process should be performed using different components of the training and testing datasets. If the database is relatively large, a sensitivity test can be performed to replace the cross-validation, in order to avoid a high computational cost[41].

It should be noted that, for a typical ANN algorithm (e.g., BPNN), it is necessary to optimize the overall ANN structure before deciding the final numbers of the hidden layer and hidden nodes. Repeated training and testing should be performed with different ANN structures. On the one hand, if the numbers of hidden layers and/or hidden neurons are too high, there is a risk of over-fitting; on the other hand, if their numbers are too low, this leads to under-fitting. Usually, the best ANN algorithmic configuration can be defined by comparing the average RMSEs from the testing sets during a cross-validation or sensitivity test[42].

## 3 Applications of ANN for Catalysis: Experiment

Generally speaking, the comprehensive evaluation indexes of industrial catalysts are activity, selectivity and service life, and the catalyst for specific reaction may also have good heat resistance, mechanical strength and anti-carbon properties, so the development cycle is often very long. When it comes to specific reactions, the preparation methods and operating conditions of catalysts depend more on practical experience. In order to further study the micro-reaction mechanism of catalysts, there are often multiple mechanism models for the same reaction, but these models often have their limitations. However, there are many factors affecting the properties of the catalyst. On the one hand, there is interaction between its own properties, such as active metal, preparation method, preparation conditions and activation conditions, and on the other hand, reaction conditions will also affect its activity. If all the experimental verification will consume huge energy. The nonlinear mapping ability of BP neural network can achieve certain precision prediction with less data in a short period of time, which is particularly efficient.

### 3.1 Prediction of Catalytic Activity

One of the earliest works for the catalytic application was done by Kito et al. in 1994[43], and they predicted the product distribution of ethylbenzene oxidative hydrogenation, with the product components of styrene, benzaldehyde, benzene + toluene, CO, and $CO_2$ as the outputs of a network. In terms of the inputs of the ANN, they used nine different independent variables that had potential relationships with the productivity and selectivity of the catalytic reaction, including: unusual valence, surface area of the catalyst, amount of catalyst, typical valence, ionic radius, coordination number, electronegativity, partial charge of oxygen ion, and standard heat of formation of oxides. Their results found that with a good experimental database, a single hidden layer ANN could perform precise predictions for the product selectivity. Sasaki et al.[44] first proposed that ANNs could be used for catalytic activity predictions, as well as experimental condition optimizations. Setting the experimental conditions such as compositional quantity, temperature, and pressure, they showed that the yield and byproducts of NO decomposition over the Cu/ZSM-5 zeolite catalyst could be precisely predicted by a well-trained ANN. For other more complicated reactions, such as 1-hexene epoxidation catalyzed by polymer supported Mo(VI) complexes, Mohammed et al. [45] showed that the ANN had a powerful predictive capacity for forecasting its catalytic activities, in excellent agreement with their experimental conclusions. For photocatalysis, the ANN also shows its strong predictive capacity. Frontistis et al. [46] studied the photocatalytic degradation of 17-ethynylestradiol (EE2) using $TiO_2$ catalysts with varying concentrations. With the inputs of reaction time, TiO2 concentration, EE2 initial concentration, matrix dissolved organic carbon (DOC), and matrix conductivity as the inputs, they found that a single hidden layer ANN could perform the minimized average RMSE during the testing processes. In terms of the biotechnical catalysis, Rahman et al.[47] also found that using the temperature, reaction time, substrate molar ratio, and enzyme amount as the inputs, an optimized ANN structure could be used for the yield prediction of lipase-catalyzed dioctyl adipate synthesis. Recently, with the developing concept of data mining, Gunay and Yildirim[48] successfully used 1337 data points from 20 studies of selective CO oxidation over Cu-based catalysts. They concluded that ANN modeling could be used to extract valuable experimental results from previous literature data and provides powerful guidance for future experimental designs. In addition to catalysis, Raccuglia et al.[49] further found that a similar concept could even help assist the materials discovery from failed experimental data. From these typical studies, it can be seen that with different types of reaction systems, catalysis, and datasets, the optimal ANN structures for prediction are significantly different. As we can see, different reaction types have very different input variables and output indicators. This means that each prediction task should be predicted by a specific model with an optimal weight contribution and network structure.

### 3.2 Optimization of Catalysis

In addition to the activity prediction, scientists started to think about a more practical question: how can we cost-effectively design novel catalysts using the predictive power of ANN? Now that we know that ANNs can precisely predict the catalytic performances of various catalytic systems, we may want to design and generate new inputs of new expected catalysts, and acquire their predicted performances from a well-trained ANN. A general algorithmic flow chart of catalyst optimization summarized by Maldonado and Rothenberg is reconstructed in Figure 2.

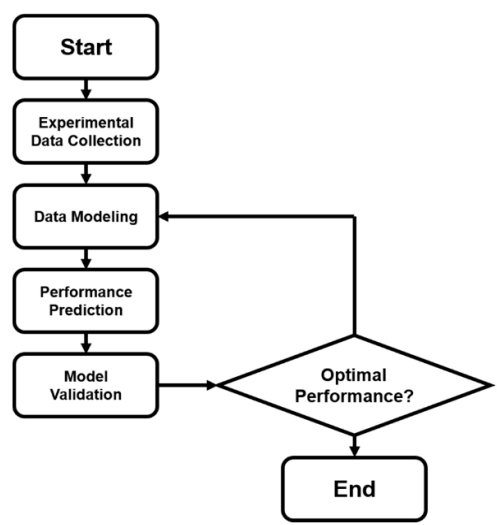

**Figure 2.** Optimization modeling that combines both experimental and computational steps.

This is actually more challenging than a straightforward prediction. For modeling and optimizing the catalysis, one of the primary works was done by Corma et al.[50], who first applied ANNs for the optimization of potential catalyst compositions for the oxidative dehydrogenation of ethane (ODHE). It should be noted that in this catalysis optimization, a genetic algorithm (GA) was introduced as the promoter of design generation. Omata and Yamada[51] developed an ANN to predict the effective additives to a Ni/active carbon (AC) catalyst for methanol vapor-phase carbonylation. Using a well-trained network, they found that Sn was an effective element that could optimize this catalytic process. Hou et al.[52] first proposed an ANN-based computer-aided framework for catalyst design. They found that such a method could be effective for the design of promising propane ammoxidation catalysts. In a similar catalytic system for propane ammoxidation, Cundari et al.[53] further combined the ANN with a GA method for a quick catalyst selection. With the use of the GA method, people can more rationally design the catalysts by optimizing the inputs of ANNs. Similarly, Umegaki et al.[54] combined the GA and ANN with a parallel activity test for optimizing a Cu-Zn-Al-Sc oxide catalyst for methanol synthesis. Rodemerck et al.[55] generalized the GA-assisted ANN method and proposed a general framework for new solid catalytic materials screening, in good agreement with their experimental data. Based on the previous developments of the GA-assisted ANN methods, Baumes et al.[56] further developed an "ANN filter" for the high-throughput screening (HTS) of heterogeneous catalysis discovery. Using the water-gas shift (WGS) reaction as an example, they showed that though the optimization method previously developed by Corma et al.[50] was successful for ODHE (as mentioned at the beginning of this Subsection), it failed to precisely estimate the WGS reaction activities. However, with a well-trained ANN classifier as a filter that could help define the "good" and "bad" catalysts, WGS catalysts could be rationally designed with a GA-assisted HTS method.

Based on the BP neural network, Baroi et al.[57] established a correlation model for the structural properties of the supported H-Y zeolite catalyst, such as the micropore area, mesopore area, pore size and loading amount, and the esterification reaction activity. Only 8 models were simulated and the reliable results were obtained. Wu et al. used momentum factor-adaptive learning rate to improve BP neural network model and orthogonal design to adjust the process parameters of high-energy ball milling WC-MgO to control its grain size., BP model through item by item, intensive scanning techniques, diameter of grinding ball, ball mill speed and ball material ratio in the operating range value, ensure the comprehensive forecast samples. The experimental verification results show that the optimal catalyst particle size training times designed by the BP improved model with the optimal structure is only 38 times, and the prediction accuracy is also better than that of the orthogonal design. Huang Kai et al. improved BP neural network by using L-M algorithm and matrix algorithm respectively, and proposed a hybrid model based on improved BP and genetic algorithm to optimize $Fe_3O_4$ composite oxide catalyst to improve methane hydrogen production performance. In this model, the catalyst life and hydrogen generation rate are composed of target parameters, and GA algorithm is used for global optimization, which improves the convergence rate, accuracy and generalization ability of the network. The hydrogen generation rate of catalyst with auxiliary element ratio and preparation conditions is higher than that of the same experimental conditions, and its lifetime is extended by up to 150%. Similarly, Abbasi et al. also designed nano-modified perovskite catalysts using a mixture model of GA and BP (Ann-GA), and predicted the reaction performance of $CH_4$-$CO_2$ with only 20 groups of experimental data for the metal molar ratio of the catalyst. The predicted value of catalyst designed by the global optimization of genetic algorithm is in good agreement. Hadi *et al.* [58] used AN-GA model and RSM to design bimetallic catalyst for MTP reaction, the propylene selectivity of the catalyst prepared by the former is higher than that of the latter in terms of the Ce load, calcination temperature and calcination time, and the prediction accuracy is also satisfactory.

## 4  Conclusion and Prospects

As the most straightforward application, ANN has been widely used for the prediction of catalytic performance during the past two decades. In the catalysis community, the optimization and design of catalysts are usually more important. In addition to predicting the catalytic activities, some studies generated new input combinations for a well-trained ANN model, and acquired the predicted output activities. In terms of the theoretical catalysis study, ANN has proven to be a good tool for catalytic descriptor prediction.

Though there is a huge success of ANNs that could facilitate the progress of the catalysis community, there are still some common challenges that should be addressed in future studies. For example, most of the relevant studies have been done by a conventional ANN (e.g., BPNN), However, with the development of machine learning, conventional ANNs are sometimes no longer the best choice. Compared to other research areas, the applications of ANN for the catalysis community are still not popular and not well-studied.